\documentclass[11pt,document,nofootinbib,superscriptaddress,onecolumn,preprintnumbers,balancelastpage]{article}
\pdfoutput=1

\usepackage{jheppub}
\usepackage{graphicx}
\usepackage{epstopdf}
\usepackage{dcolumn}% Align table columns on decimal point
\usepackage{bm}% bold math
\usepackage{hyperref}
\usepackage{color}
\usepackage{amsmath}
\usepackage{caption}
\usepackage{booktabs}
\def\beq{\begin{align}}
\def\eeq{\end{align}}
\newcommand{\gsim}{ \mathop{}_{\textstyle \sim}^{\textstyle >} }
\newcommand{\lsim}{ \mathop{}_{\textstyle \sim}^{\textstyle <} }

\def\mpl{M_{\rm Pl}}

\title{
QCD Axion Dark Matter from a Late Time Phase Transition
}

\author[1]{Keisuke Harigaya}
\author[2,3]{Jacob M.~Leedom}
\affiliation[1]{School of Natural Sciences, Institute for Advanced Study, Princeton, New Jersey, 08540}
\affiliation[2]{Department of Physics, University of California, Berkeley, California 94720, USA}
\affiliation[3]{Theoretical Physics Group, Lawrence Berkeley National Laboratory, Berkeley, California 94720, USA}

\abstract{
We investigate the possibility that the Peccei-Quinn phase transition occurs at a temperature far below the symmetry breaking scale. Low phase transition temperatures are typical in supersymmetric theories, where symmetry breaking fields have small masses. We find that QCD axions are abundantly produced just after the phase transition. The observed dark matter abundance is reproduced even if the decay constant is much lower than $10^{11}$ GeV. The produced axions tend to be warm. For some range of the decay constant, the effect of the predicted warmness on structure formation can be confirmed by future observations of 21 cm lines. A portion of parameter space requires a mixing between the Peccei-Quinn symmetry breaking field and the Standard Model Higgs, and predicts an observable rate of rare Kaon decays.
}

\date{\today}

\begin{document}

\maketitle

\tableofcontents

%%%%%%%%%%%%%%%%%%%
%---------------SECTION-------------------%
%%%%%%%%%%%%%%%%%%%
\section{Introduction}
The CP violation in QCD~\cite{tHooft:1976rip}, expressed by the so-called $\theta$ parameter, is extremely small -  $\theta < 10^{-10}$~\cite{Baker:2006ts}. The smallness of the CP violation is elegantly explained by the Peccei-Quinn (PQ) mechanism~\cite{Peccei:1977hh,Peccei:1977ur}. One introduces a spontaneously broken global symmetry which is explicitly broken by the QCD anomaly, and predicts a pseudo-Nambu-Goldstone boson called an axion~\cite{Weinberg:1977ma,Wilczek:1977pj}. For a large enough symmetry breaking scale, the axion is stable and a dark matter candidate~\cite{Preskill:1982cy,Abbott:1982af,Dine:1982ah}.

Two production mechanisms of axion dark matter in the early universe are widely recognized. One is the misalignment mechanism~\cite{Preskill:1982cy,Abbott:1982af,Dine:1982ah}, where the displacement of the axion field from the vacuum turns into oscillations which behave as dark matter. Another is the emission of axions from the string-domain wall network produced after the spontaneous breaking of the PQ symmetry~\cite{Davis:1986xc,Kawasaki:2014sqa,Klaer:2017ond,Buschmann:2019icd}. Both mechanisms require that the axion decay constant $f_a$ is large -  $f_a \gsim 10^{11}$~{\rm GeV}.
(The estimation of the abundance in the latter mechanism assumes a scaling law of topological defects. See~\cite{Gorghetto:2018myk,Kawasaki:2018bzv,Martins:2018dqg} for a possible violation as well as~\cite{Hindmarsh:2019csc} for results that contradict this violation.)

In this paper, we point out a new production mechanism for axion dark matter under the assumption that the phase transition temperature of the PQ symmetry breaking is far below the symmetry breaking scale. We find that axions are abundantly produced via parametric resonance arising from oscillations of the symmetry breaking field~\cite{Traschen:1990sw,Kofman:1994rk,Shtanov:1994ce,Kofman:1997yn} after the phase transition.
Since the phase transition temperature is low, the produced axions are not thermalized and remain as dark matter.
The axions produced from the late time phase transition can explain the observed dark matter abundance even if the decay constant is much smaller than $10^{11}$~GeV.

Low phase transition temperatures are natural in supersymmetric theories. This is because the radial direction of the PQ symmetry breaking field, commonly called the saxion, is the scalar partner of the nearly-massless axion. The mass of the saxion is given by a supersymmetry breaking soft mass and is much smaller than the PQ symmetry breaking scale. This small mass in turn yields a relatively low phase transition temperature.

In contrast to the two conventional mechanisms, the produced axions are initially relativistic and red-shift sufficiently to be dark matter. In some of the parameter space, the axions are still warm enough to affect structure formation by an observable amount.

There are intensive ongoing and future experimental efforts to search for the axion with a small decay constant, such as  IAXO~\cite{Vogel:2013bta,Armengaud:2014gea}, TASTE~\cite{Anastassopoulos:2017kag}, Orpheus~\cite{Rybka:2014cya}, MADMAX~\cite{TheMADMAXWorkingGroup:2016hpc}, ARIADNE~\cite{Arvanitaki:2014dfa,Geraci:2017bmq}, and many others~\cite{Sikivie:2014lha,Arvanitaki:2017nhi,Baryakhtar:2018doz,Lawson:2019brd}. Astrophysical observations and the above searches for the QCD axion could probe the dynamics of the PQ phase transition.

%%%%%%%%%%%%%%%%%%%
%---------------SECTION-------------------%
%%%%%%%%%%%%%%%%%%%
\section{Late time PQ phase transition}
%%%%%%%%%%%%%
%------SUBSECTION-------%
%%%%%%%%%%%%%
\subsection{The Model}
We consider a coupling of the PQ symmetry breaking field $P$ to new PQ charged fermions $\psi$ and $\bar{\psi}$,
\begin{align}
\label{eq:yuk}
{\cal L} = y P \psi \bar{\psi} + \text{h.c.}.
\end{align}
The new fermions may be identified with the hidden quarks of the KSVZ model~\cite{Kim:1979if,Shifman:1979if}.
This coupling gives a thermal potential to $P$,
\begin{align}
V_T = V(P) + V_{\rm th}(P,T),
%V = (y^2 T^2 - m^2)|P|^2,
\end{align}
where $V$ is the vacuum potential of $P$. In a typical second-order phase transition, the thermal potential can be expanded about $\lvert P\rvert/T \ll 1$ to get the thermal mass and higher corrections:
\begin{align}
V_{\rm th}(P,T) = y^2 T^2|P|^2 + \cdots.
\end{align}
The critical temperature $T_c$ is then defined as the temperature at which the curvature of the potential about the origin vanishes. If the vacuum potential has the form $V = -m^2\lvert P\rvert^2 + \cdots$,
\begin{align}
\label{eq:Tc}
T_c = \frac{m}{y},
\end{align}
We define a late time phase transition as a phase transition that satisfies $m\ll f_a$ so that $T_c \ll f_a$.

While a late time phase transition may seem fine-tuned, the above hierarchy is typically encountered in supersymmetric theories where PQ symmetry breaking scales much larger than the mass $m$ are naturally obtained through the stabilization of $P$ by a higher dimensional interaction~\cite{Murayama:1992dj}, 
\begin{align}
\label{eq:V1}
V = \left(\frac{2^{n-2}m_s^2 }{n(n-1) f_a^{2n-2}  }\right) \lvert P\rvert^{2n} - \frac{m_s^2}{2n-2}\lvert P\rvert^2 + \frac{m_s^2f_a^2}{4n}
\end{align}	
with $n>2$,
or through the renormalization group running of the soft mass of $P$~\cite{Moxhay:1984am},
\begin{align}
\label{eq:V2}
V = \frac{m_s^2}{2} |P|^2 \left( {\rm ln}\frac{2|P|^2}{f_a^2} -1  \right) + \frac{1}{4}m_s^2 f_a^2.
\end{align}
The parameters of these potentials have been chosen such that the saxion mass around the vacuum is $m_s~(\sim m)$, the vacuum expectation value of $|P|$ is $f_a/\sqrt{2}$, and the vacuum energy at the minimum of the potential vanishes.

\subsection{The phase transition and thermal inflation}

%We note that the transition rate from the false vacuum $\langle P\rangle = 0$ to the true vacuum at a given temperature is negligible~\cite{Yamamoto:1985rd,Hiramatsu:2014uta}, and so the false vacuum is stable enough that thermal inflation does occur.

\begin{figure}[!tb]
\centering
\includegraphics[width=0.7\textwidth]{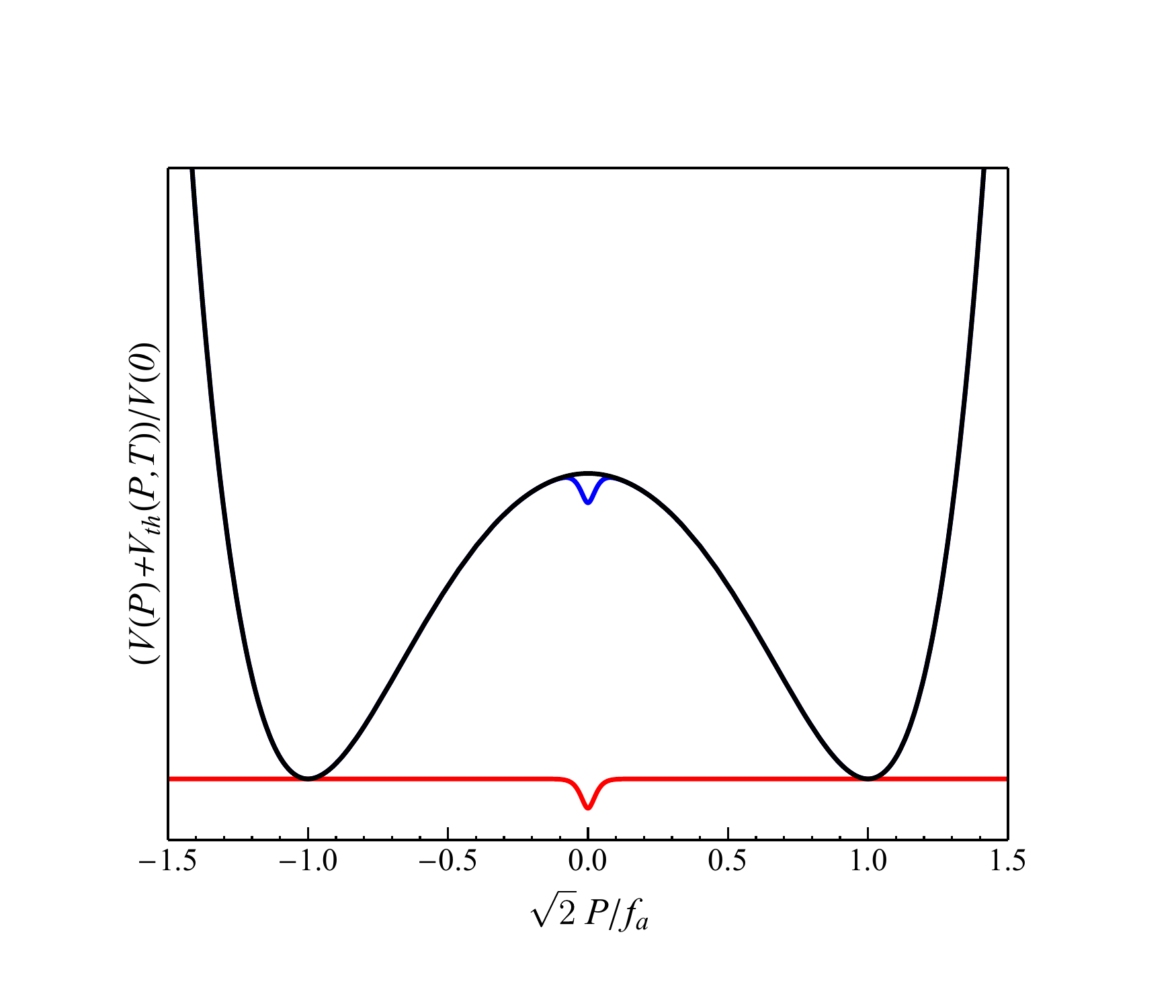}
\caption{The solid black line is the vacuum potential of $P$ as given in Eq.~(\ref{eq:V1}) with $n=3$. The solid red line is the thermal correction to the vacuum potential that has been exaggerated to emphasize the structure. The blue line is the sum of the vacuum potential and exaggerated thermal potential. }
\label{vth}
\end{figure}

A late time phase transition does not proceed via a first- or second-order phase transition. Since $m_s \ll f_a$, the high temprature expansion is insufficient at $T_c$ and one must follow the evolution of the total thermal potential as the temperature decreases.  At high temperatures, $T^4 \gg m_s^2 f_a^2$, the origin is an absolute minimum. For  $T^4 < m_s^2 f_a^2$, on the other hand, the minimum at the origin is a local, metastable minimum until $T_c$. The scalar field $P$ is trapped at the origin by the dip formed by the thermal correction. This is schematically displayed in Fig.~\ref{vth}. One might assume that the phase transition is then first order and proceeds via bubble nucleation. However, the numerical results of~\cite{Hiramatsu:2014uta} find that a slightly different process occurs. Before the quantum tunneling rate becomes effective enough for bubble percolation, thermal fluctuations of $P$ are large enough to cause the phase transition. For a weak Yukawa coupling, $y \lesssim 0.1$, this is found to occur for a temperature within a sub percent of $T_c$~\cite{Hiramatsu:2014uta}.

For both of the potentials above, the PQ symmetry breaking field at the origin has a potential energy density $V(0) \propto m_s^2 f_a^2$. This is larger than the radiation energy density at $T_c$, $\rho_{rad} = \frac{\pi^2}{30}g_*T_c^4$,  if 
\begin{align}
y \gtrsim \sqrt{\frac{m_s}{f_a}},
\end{align}
which we assume in the following. The case with smaller $y$ can be analyzed in a similar manner. Since the potential energy dominates, a period of so-called thermal inflation~\cite{Yamamoto:1985rd,Lyth:1995ka} occurs with a Hubble scale $H_{\rm PT}\propto \frac{m_s f_a}{\mpl}$ before the phase transition.

%%%%%%%%%%%%%
%------SUBSECTION-------%
%%%%%%%%%%%%%
\subsection{Axions from inhomogeneity}
Just after the phase transition, the configuration of the PQ symmetry breaking field is inhomogeneous. Under a normal second-order phase transition, the correlation length of the configuration is determined by the Kibble-Zurek mechanism~\cite{Kibble:1976sj,Zurek:1985qw}. However, our scenario differs from this mechanism. As discussed above, \cite{Hiramatsu:2014uta} finds that the phase transition occurs by thermal fluctuations at a temperature $T > T_c$. %If the Yukawa coupling introduced in Eq.~(\ref{eq:yuk})  is weak, $y \lesssim 0.1$, then the phase transition will occur at a temperature that is within a few per cent of $T_c$ so that the correlation length of $P$ around the phase transition is much larger than $m_s^{-1}$.
 In order to estimate the correlation length of the inhomogeneous configuration, we assume the weak coupling scenario of~\cite{Hiramatsu:2014uta} so that the phase transition occurs at $T_s = \alpha T_c$, where $0 < \alpha-1 < 10^{-2}$. The correlation length of the scalar field $P$ at $T_s$ is
\begin{align}
\xi_s = \frac{1}{m_s}\bigg(\frac{\lvert T_s-T_c\rvert}{T_c}\bigg)^{-1/2} = m_s^{-1}\lvert 1-\alpha\rvert^{-1/2}
\end{align}
Since the field is correlated on the length scale $\xi_s$, we can draw an analogy with the Kibble-Zurek mechanism and expect typically one cosmic string per correlation length volume, $\xi_s^3$, with energy density $f_a^2 / \xi_s^2$.
The gradient energy density of the inhomogeneity is $f_a^2 / \xi_s^2$. Typically one cosmic string per correlation length volume, $\xi_s^3$, exists with energy density $f_a^2 / \xi_s^2$.

The inhomogeneous configuration is quickly homogenized until the correlation length becomes as large as the horizon size. The gradient and string energy should be emitted as axions with typical wavelength $\xi_s$. The number density of axions produced  from the inhomogeneity is then
\begin{align}
		n_a^{\rm inh} \sim \frac{f_a^2}{\xi_s} = m_sf_a^2 \lvert 1-\alpha			\rvert^{1/2}.
\end{align}
The potential energy density $m_s^2 f_a^2$ is converted into the oscillation energy of the saxion, which subsequently red-shifts in proportion to the inverse cube of the scale factor of the universe.
We thus normalize the number density of axions by the energy density of the saxion oscillation,
${n_a}/{\rho_s}$,
which does not change under red-shifting. For axions coming from the inhomogeneity,
\begin{align}
\label{eq:str_rat}
\frac{n_a^{\rm inh}}{\rho_s} = \frac{1}{m_s}\lvert 1-\alpha \rvert^{1/2}.
\end{align}
As we will see in the next section, however, axions produced from the inhomogeneity are subdominant.

%%%%%%%%%%%%%%%%%%%
%---------------SECTION-------------------%
%%%%%%%%%%%%%%%%%%%
\section{Axions from parametric resonance}
%%%%%%%%%%%%%
%------SUBSECTION-------%
%%%%%%%%%%%%%
\subsection{Axion Production}
The axion population produced by cosmic strings is supplemented and surpassed by a second production mechanism - parametric resonance~\cite{Traschen:1990sw,Kofman:1994rk,Shtanov:1994ce,Kofman:1997yn}. After the phase transition, the saxion oscillates with an amplitude on the order of $f_a$ and induces a time-dependent dispersion relation in the equation of motion for axion modes. In certain momentum bands, the axion mode solutions feature instabilties that grow exponentially in time. These modes then yield the axion population produced by the non-perturbative process of parametric resonance.

The production rate of axions via parametric resonance is as large as the frequency of the oscillations $\sim m_s$, since that is the only energy scale appearing in the equation of motion of axions. This is explicitly shown in Appendix A, where we display that the rate of axion production is $m_s$ times an $\mathcal{O}(1)$ constant (Fig.~\ref{fig:mu}). By comparing this axion production rate with the Hubble rate $ \sim m_s (f_a /\mpl)  \ll m_s$, we see that the parametric resonance process is very efficient. 

Parametric resonance preferentially creates axions with momenta $k_a \sim m_s / 2$ and continues until the newly produced axion energy density is roughly equal to the initial saxion energy density, which is just the potential energy at the origin $V(0) \propto m_s^2f^2_a$. We label this second contribution to the axion density as $n_a^{\rm PR}$, so that
\begin{align}
\label{eq:pr}
\frac{n_a^{\rm PR}}{\rho_s} \simeq \frac{m_s^2f_a^2}{m_s}\frac{1}{m_s^2f_a^2}= \frac{1}{ m_s}.
\end{align}
For saxion oscillations with an amplitude of the order of $f_a$, saxion fluctuations are also produced by parametric resonance and obtain a number density similar to the one in Eq.~(\ref{eq:pr}).

Comparing Eqs.~(\ref{eq:str_rat}) and (\ref{eq:pr}), we see that the parametric resonance axions are the dominant contribution to the axion population. In what follows, we only take into account the parametric resonance axions.
With this, one finds that the axion number density normalized by the entropy density $s$ is
\begin{align}
 Y_a = \frac{n_a}{s} = \frac{n_a}{\rho_s}\frac{\rho_s}{s} \simeq \frac{T_{RH}}{m_s},
\end{align}
where $T_{\rm RH}$ is the reheat temperature by the dissipation of the saxion oscillation and fluctuations after the thermal inflation.
To obtain the axion dark matter abundance
\begin{align}
			Y_a^{\rm DM} =\frac{1}{m_a}\frac{\rho_{DM}}{s} = 70\left(\frac{f_a}{10^9 \text{ GeV}}		\right),
\end{align}
the reheat temperature $T_{\rm RH}$ must be above
\begin{align}
\label{eq:TDM}
T_{\rm DM} \simeq 0.7 \text{ GeV} \left(\frac{m_s}{10 \text{ MeV}}\right)\left(\frac{f_a}{10^9 \text{ GeV}}\right).
\end{align}
If $T_{\rm RH}$ is higher than this value, axions are overproduced, but the introduction of extra entropy production from heavy fields can generate the correct abundance. 
Obtaining this reheat temperature is discussed below.
%%%%%%%%%%%%%
%------SUBSECTION-------%
%%%%%%%%%%%%%
\subsection{Validity of Parametric Resonance Scenario}

One might be concerned that the inhomogeneity caused by the phase transition could ruin the parametric resonance process, which requires coherent oscillations. We do not anticipate that this is the case since the wavelengths in the resonance band are $\sim 1/m_s$, which is much shorter than the length scale on which the PQ symmetry breaking field is correlated, $\xi_s$. Hence the oscillations are effectively coherent for the modes in the resonance band.

The above scenario could also be affected if energy from the saxion oscillations is drained into Standard Model fields. This is only a concern if the rate of energy loss to Standard Model fields is comparable to $m_s$, the rate of axion production from parametric resonance. The saxion coupling to gluons gives a thermalization rate~\cite{Mukaida:2012qn,Bodeker:2006ij}
\begin{align}
\frac{\Gamma_{\rm gluon}}{m_s} = \bigg(A \frac{\alpha_3^2T^3}{f_a^2}\bigg)\frac{1}{m_s} \lesssim A\alpha_3^2 \sqrt{\frac{m_s}{f_a}} \ll 1,
\end{align}
where A is an $\mathcal{O}(10^{-3})$ constant and we used an inequality $T^4 \lesssim m_s^2f_a^2$. Thus the energy loss to the Standard Model through gluons is negligible during parametric resonance. Non-perturbative production of gluons is ineffective due to the loop-suppressed coupling between gluons and saxions. The thermal mass of the gluon further reduces the effectiveness of non-perturbative production.

In the sequel, we consider a saxion-Higgs mixing that provides a second avenue for energy loss to Standard Model fields. The ratio of the thermalization rate from the coupling $\lambda S^2H^\dag H$~\cite{Mukaida:2012qn} to $m_s$ is 
\begin{align}
\label{eq:higtherm}
\frac{\Gamma_{\rm Higgs}}{m_s}= \bigg(\frac{\lambda^2 f_a^2}{T}\bigg)\frac{1}{m_s} \lesssim \frac{10^{-8}m_H^4}{f_a^{1/2}m_s^{3/2}v_{\rm EW}^2},
\end{align}
where $m_H = 125$ GeV and $v_{\rm EW} = 246$ GeV. We have used the experimental upper bound $\theta \lesssim 10^{-4}$~\cite{Beacham:2019nyx} on the mixing angle $\theta \sim \frac{\lambda f_av_{EW}}{m_H^2}$. The last quantity in Eq.~(\ref{eq:higtherm})  is much smaller than unity in the viable space discussed below and we conclude that this process is also ineffective. Higgs particles can be produced by parametric resonance, but the produced Higgses are immediately dissipated by decays or scatterings with the thermal bath, and so the Bose-enhancement necessary for efficient parametric resonance is absent. From these results we see that our scenario of parametric resonance is not spoiled by interactions with the Standard Model.

%%%%%%%%%%%%%%%%%%%
%---------------SECTION-------------------%
%%%%%%%%%%%%%%%%%%%
\section{Model Constraints}
\subsection{Axion Warmness}
The axions are produced relativistically and may behave as warm dark matter. To be concrete, we consider a specific model with the PQ potential in Eqs.~(\ref{eq:V1}) or (\ref{eq:V2}).
The expressions in the following depend on the potential, but a similar analysis can be performed for more general potentials.

\begin{figure}[!tbh]
\centering
\includegraphics[clip, trim=0.5cm 1.3cm 0.5cm 2.3cm, width=0.9\textwidth]{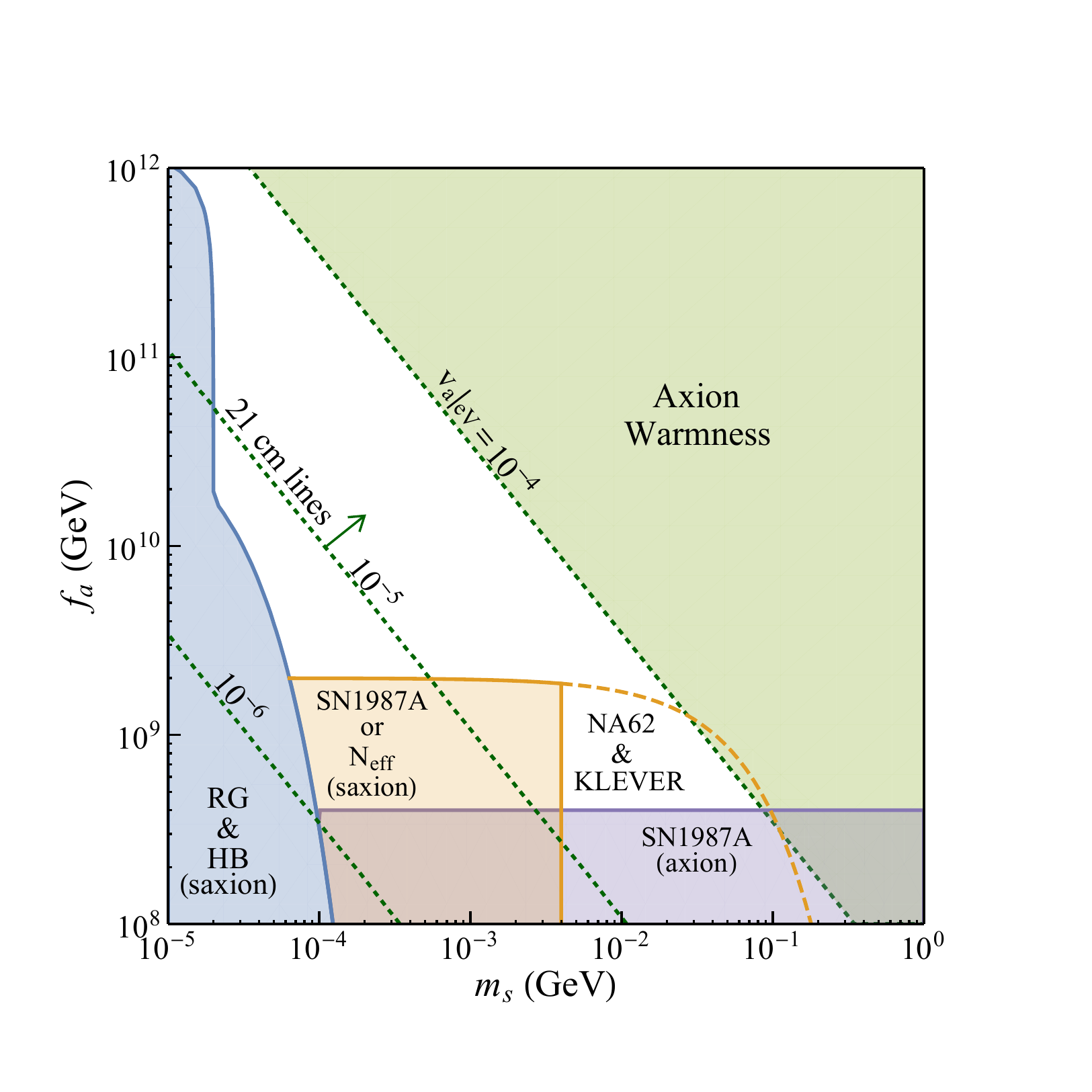}
\caption{Constraints on the saxion mass $m_s$ and the axion decay constant $f_a$.}
\label{fig}
\end{figure}

To estimate the warmness, we first note that the ratio between the axion momentum, $k_a$, and the cube root of the axion number density, $n_a^{1/3}$, is constant throughout the evolution of the universe, 
\begin{align}
		\frac{k_a}{n_a^{1/3}} = \left(\frac{n}{2}\right)^{\frac{1}{3}}\left(\frac{m_s}{f_a}\right)^{\frac{2}{3}},
\end{align}
where $n$ is an integer larger than $2$ in Eq.~(\ref{eq:V1}) or $1$ for the potential in Eq.~(\ref{eq:V2}).
Here it is assumed that half of the potential energy of the saxion is transferred into axions with momenta $k_a = m_s/2$. 
Using the observed dark matter abundance to fix $n_a$ relative to the entropy density $s$, we obtain
\begin{align}
			v_a \simeq 6\times 10^{-4}n^{1/3}\left(\frac{f_a}{10^9 \text{ GeV}}\right)^{\frac{2}{3}}\left(\frac{m_s}{{\rm GeV}}\right)^{\frac{2}{3}}\left(\frac{T}{ \text{eV}}\right)
		\end{align}
for the velocity of the axions at temperature $T$. We have assumed $T \ll $ MeV to express the entropy density in terms of $T$. In Fig.~\ref{fig}, we show contours of the axion velocity at $T = 1$ eV for $n=3$.

The constraint on the warmness of dark matter is frequently estimated for a model where dark matter consists of a massive Weyl fermion with mass $m_{\rm WDM}$ that decouples while relativistic and is later diluted.
In such a model, the typical velocity of dark matter, $v_{\rm WDM}$, at temperature $T$ is given by
\begin{align}
v_ {\rm WDM} = \frac{k_{\rm WDM}}{m_{\rm WDM}} \simeq 10^{-4} \left(\frac{T}{1\text{ eV}}\right)\left(\frac{3.3\text{ keV}}{m_{\rm WDM}}\right)^{\frac{4}{3}}.
\end{align}
This result, combined with the warm dark matter mass bound, $m_{\rm WDM}>3.3$ keV~\cite{Viel:2013apy}, yields the generic velocity bound of $v < 10^{-4} $ at $T= 1$ eV.
This warmness bound imposes the following constraint on the saxion mass
\begin{align}
			m_s \lsim 30 ~{\rm MeV} \left( \frac{3}{n} \right)^{\frac{1}{2}}\left(\frac{10^9 \text{ GeV}}{f_a}\right).
\end{align}
The green shaded region in Fig.~\ref{fig} is disfavored by this constraint.
Future observations of 21cm lines can probe $m_{\rm WDM} < 10\mathchar`-20$ keV~\cite{Sitwell:2013fpa}, which corresponds to $v_a \gsim 10^{-5}$ at $T =1$~eV, as indicated by the arrow in Fig.~\ref{fig}. 
For convenience, we provide the correspondence between the mass of this fermionic dark matter and the parameters of our model,
\begin{align}
		m_{\rm WDM} \leftrightarrow \frac{0.8 ~{\rm keV}}{n^{1/4}} \left(\frac{10^9 \text{ GeV}}{f_a}\right)^{\frac{1}{2}}\left(\frac{\rm GeV}{m_s}\right)^{\frac{1}{2}}.
\end{align}

%%%%%%%%%%%%%
%------SUBSECTION-------%
%%%%%%%%%%%%%
\subsection{Stellar Cooling}
In addition to the warmness bound, we consider constraints from the cooling of red giant (RG) and horizontal branch (HB) stars~\cite{Raffelt:1996wa,Grifols:1986fc,Grifols:1988fv} by the emission of saxions. We follow the analysis performed in~\cite{Hardy:2016kme, Knapen:2017xzo}. For RG and HB stars, one must demand that the energy transport by new particles with effective nucleon couplings not exceed 10 erg g$^{-1}$s$^{-1}$. These constraints are displayed as the blue region in Fig.~\ref{fig} labeled as ``RG \& HB".

%%%%%%%%%%%%%
%------SUBSECTION-------%
%%%%%%%%%%%%%
\subsection{SN1987A and  N$_{\rm eff}$}
The orange shaded and dashed excluded parameter regions in Fig.~\ref{fig} labeled ``SN1987A or N$_{\rm eff}$" arise from the SN1987A constraint of~\cite{Ishizuka:1989ts}, as well as the constraint on the effective number of relativistic degrees of freedom $N_{\rm eff}$~\cite{Aghanim:2018eyx}. For SN1987A, the energy loss should not exceed 10$^{19}$ erg g$^{-1}$s$^{-1}$. This leads to the boundary of the orange shaded region as well as the orange dashed curve. If one takes the SN1987A constraint on energy loss directly, the region below the orange dashed curve would be excluded. However, if one assumes a coupling between the saxion and Standard Model Higgs of the form $ \lambda S^2H^\dag H$, then the saxion enters the so-called trapping regime and the SN1987A constraint does not apply and the region below the dashed orange curve is permitted. To be in the traping regime, the saxion-Higgs mixing angle $\theta \sim \frac{\lambda f_av_{EW}}{m_H^2}$ must be larger than $\sim 10^{-4.5}$~\cite{Beacham:2019nyx}. The orange shaded region remains constrained since the mixing keeps the saxion in thermal equilibrium with electrons even after neutrinos decouple in the early universe. Hence the depletion of the saxion energy heats up photons, resulting in $N_{\rm eff} < 3$. 
Assuming that neutrinos suddenly decouple at $T\simeq 2$ MeV, we determine a lower bound on the saxion mass of $m_s \gsim 4$ MeV. The purple shaded region is the bound related to axions arising from SN1987A~\cite{Ellis:1987pk,Raffelt:1987yt,Turner:1987by,Mayle:1987as,Raffelt:2006cw}. We also note that there is at least an order of magnitude uncertainty in the SN1987A constraints~\cite{Hanhart:2000ae,Hanhart:2000er,Rrapaj:2015wgs,Chang:2016ntp,Carenza:2019pxu,Bar:2019ifz}. This could lead to a larger parameter space. The saxion-Higgs mixing results in rare decays of Kaons. As shown in~\cite{Beacham:2019nyx}, the large mixing in the trapping regime can be probed by NA62 and KLEVER experiments~\cite{NA62:2017rwk,Ambrosino:2019qvz}.

%%%%%%%%%%%%%
%------SUBSECTION-------%
%%%%%%%%%%%%%
\subsection{Saxion Thermalization}
The saxion should be thermalized at or above the temperature $T_{\rm DM}$ in Eq.~(\ref{eq:TDM}). We consider the case where the PQ symmetry breaking field $P$ couples to a pair of new fermions $f$ and $\bar{f}$ via the Yukawa coupling
\begin{align}
\label{eq:ferm2}
{\cal L } = \frac{\mu}{f_a} P f \bar{f},
\end{align}
where $\mu$ is the mass of the fermion.
For $T > \mu$, the saxion thermalizes with a rate $\simeq 0.1 T \mu^2 / f_a^2$~\cite{BasteroGil:2010pb,Mukaida:2012qn}, leading to a reheating temperature
\begin{align}
T_{\rm RH} \simeq  100 \text{ GeV}\left(\frac{\mu}{\rm 100~{\rm GeV}}\right)^2 \left(\frac{10^9~{\rm GeV}}{f_a}\right)^2.
\end{align}
If the fermion is charged under the Standard Model gauge group, the mass $\mu$ must be above $100$ GeV. However,
This reheating temperature is larger than the lower bound given in Eq.~(\ref{eq:TDM}) above, and more than enough axion dark matter is produced. $T_{\rm RH}= T_{\rm DM}$ can be obtained through thermalization from coupling the saxion with Standard Model particles or with particles that are neutral under the Standard Model gauge group. Note that too much dark radiation is produced if the new fermions introduced in Eq.~(\ref{eq:ferm2}) have masses below $\mathcal{O}(10)$ MeV. If the fermion mass $\mu$ required to produce the dark matter abundance is below $\mathcal{O}(10)$ MeV, one must fix the fermion mass to be larger than this scale and introduce additional entropy production to dilute the overproduced axions.
%%%%%%%%%%%%%
%------SUBSECTION-------%
%%%%%%%%%%%%%
\subsection{Axion Thermalization}
Axions produced in our model are never thermalized. The thermalization rate of an axion is suppressed by the decay constant and the momentum of the axion~\cite{Moroi:2014mqa},
\begin{align}
\Gamma_a = b \frac{k_a^2}{f_a^2} T,
\end{align}
where $b$ is a constant which depends on the axion coupling. If the axion couples to gluons, $b$ is loop-suppressed and is as small as $10^{-5}$. If instead the axion couples to a light fermion in the thermal bath, $b$ may be as large as $O(1)$.
During the matter dominated era by the saxion oscillation, $k_a/ \rho_s^{1/3}$ remains constant. The momentum of axions is then given by
\begin{align}
k_a \simeq \left(\frac{m_s \rho_s}{f_a^2}\right)^{\frac{1}{3}}.
\end{align}
The energy density of the thermal bath never exceeds that of the saxion. Hence the thermalization rate is bounded from above,
\begin{align}
\Gamma_a < b \frac{m_s^{2/3} \rho_s^{11/12}}{f_a^{10/3}}.
\end{align}
The ratio between the thermalization rate and the Hubble expansion rate is
\begin{align}
\label{eq:bound_rate}
\frac{\Gamma_a}{H} <  b \frac{m_s^{2/3} \rho_s^{5/12} \mpl}{f_a^{10/3}} < b \frac{m_s^{3/2} \mpl}{f_a^{5/2}}
\end{align}
where the last inequality is saturated right after the phase transition. In the region of parameter space that produces sufficiently cold axion dark matter, the axions are never thermalized. One can see that the late-time phase transition is crucial. If the mass $m_s$ is as large as $f_a$, the thermalization is effective.
After the saxion decays and the radiation dominated era begins, the thermalization rate decreases faster than the Hubble expansion rate and the thermalization of the axions never becomes efficient.

%%%%%%%%%%%%%%%%%%%
%---------------SECTION-------------------%
%%%%%%%%%%%%%%%%%%%
\section{Discussion}
We have investigated a production mechanism for QCD axion dark matter associated with PQ symmetry breaking at a low temperature. We find that axions are primarily produced by parametric resonance via oscillations of the PQ symmetry breaking field. The low phase transition temperature fits naturally in supersymmetric theories.

The axions produced by this mechanism tend to be warm. The prediction on axion warmness is shown in Fig.~\ref{fig} and constrains the allowed parameter space. Future observations of 21cm lines will probe the parameter space further. Discovery of the QCD axion in laboratories and the determination of dark matter warmness by astrophysical observations will suggest that axion dark matter was produced by parametric resonance.

Fig.~\ref{fig} is one of the primary results of this paper and contains information beyond the warmness constraint. As outlined above, one also has bounds from energy loss in RG and HB stars and supernovae by saxion emission, as well as axion emission in the supernovae case. We note that our parameter space easily allows for rather low values of the axion decay constant, particularly if strong saxion-Higgs coupling occurs to trap saxions inside the supernova core, or if the traditional SN1987A bound is loosened. The region with large saxion-Higgs mixing can be probed by observations of rare Kaon decays.

There are several uncertainties in our estimation of the warmness. First, we have assumed that half of the energy density of the saxion oscillation is transferred into axions. In reality the transferred fraction will not be exactly half. Second, we have assumed that the momentum as well as the number density of the axions decrease only by the cosmic expansion. However, the momentum/number density can slowly increase/decrease by axion self interactions, see~\cite{Micha:2002ey} for a related discussion. These two effects will change the prediction on axion velocity by an $O(1\mathchar`-10)$ factor. Whether or not the whole parameter space can be probed depends on these uncertainties, which can be fixed by numerical computation.

We list other known mechanisms to produce axion dark matter for $f_a \ll 10^{11}$ GeV: 1)axion emission from long-lived topological defects which collapse via explicit PQ symmetry breaking~\cite{Kawasaki:2014sqa,Hiramatsu:2010yn,Hiramatsu:2012sc,Ringwald:2015dsf,Harigaya:2018ooc}, 2)parametric resonant production of axions from oscillations of the saxion with a large initial field value~\cite{Co:2017mop}, 3)a misalignment angle fine-tuned to be close to $\pi$~\cite{Turner:1985si,Lyth:1991ub,Strobl:1994wk}, 4)dynamical mechanisms that set the misalignment angle close to $\pi$~\cite{Co:2018mho,Takahashi:2019pqf}, 5)the misalignment mechanism with non-standard cosmology~\cite{Visinelli:2009kt}, and 6)delayed oscillations of the axion field because of a non-zero kinetic energy of the axion field~\cite{Co:2019wyp,Co:RM}.
Among them, 2) can also produce warm axions, but the produced axions are much colder than our mechanism for a given $(m_s,f_a)$. We also note that the large field value assumed in~\cite{Co:2017mop} requires that the potential be flat for large field values and is therefore incompatible with the potential in Eq.~(\ref{eq:V1}).

Our axion production mechanism involves a PQ symmetry breaking field that is initially trapped at the origin. We may consider a generic situation where a PQ symmetry breaking field is trapped at some other point in field space and later begins to oscillate with a large amplitude.
One example is a model with the superpotential
\begin{align}
W = \lambda X (P \bar{P} - V_{\rm PQ}^2),
\end{align}
where $P$ and $\bar{P}$ are PQ symmetry breaking fields and $X$ is a chiral multiplet that fixes them on the moduli space $P\bar{P} = V_{\rm PQ}^2$. The moduli space is lifted by additional superpotential terms that spontaneously break supersymmetry~\cite{Carpenter:2009zs,Carpenter:2009sw,Harigaya:2017dgd}, or by the soft masses of $P$ and $\bar{P}$.
Ref.~\cite{Moroi:2012vu} investigates the trapping of the PQ symmetry breaking fields on the moduli space by a thermal potential and finds that oscillations occur in some region of the parameter space. Axions should then be produced via parametric resonance in this setup as well.

\section*{Acknowledgement}
We thank David Dunsky, Masahiro Kawasaki, Simon Knapen and Toyokazu Sekiguchi for discussions. Fig.~\ref{fig} of this paper was created using the SciDraw scientific figure preparation system. This work was supported in part by the Director, Office of Science, Office of High Energy and Nuclear Physics, of the U.S.\ Department of Energy under Contract DE-AC02-05CH11231 (JL) and DE-SC0009988 (KH), by the National Science Foundation under grants PHY-1316783 (JL), and the Raymond and Beverly Sackler Foundation Fund (KH).

\appendix

\section{Parametric Resonance}

\begin{figure}[!tbh]
\centering
\includegraphics[width=0.68\textwidth]{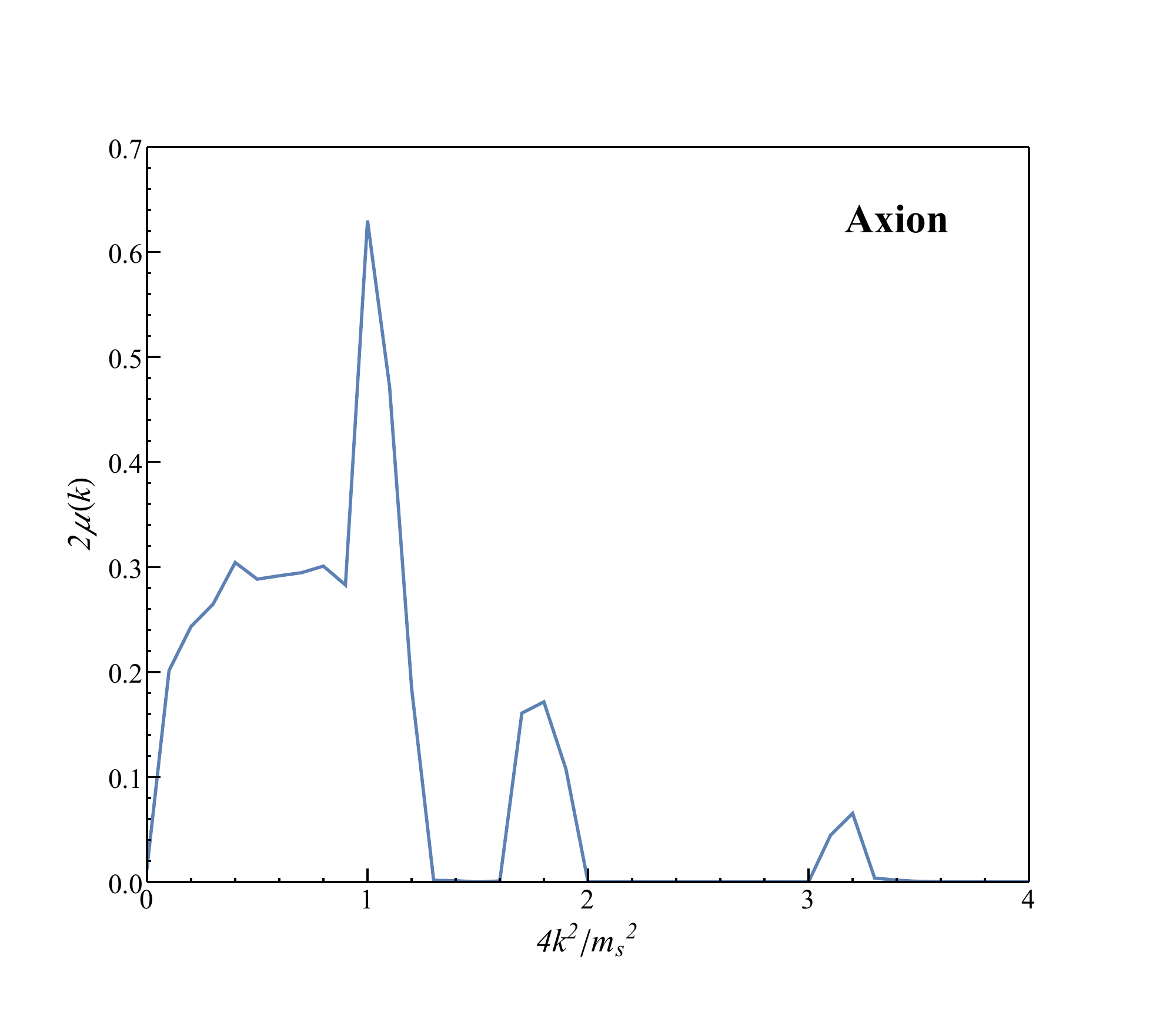}
\includegraphics[width=0.68\textwidth]{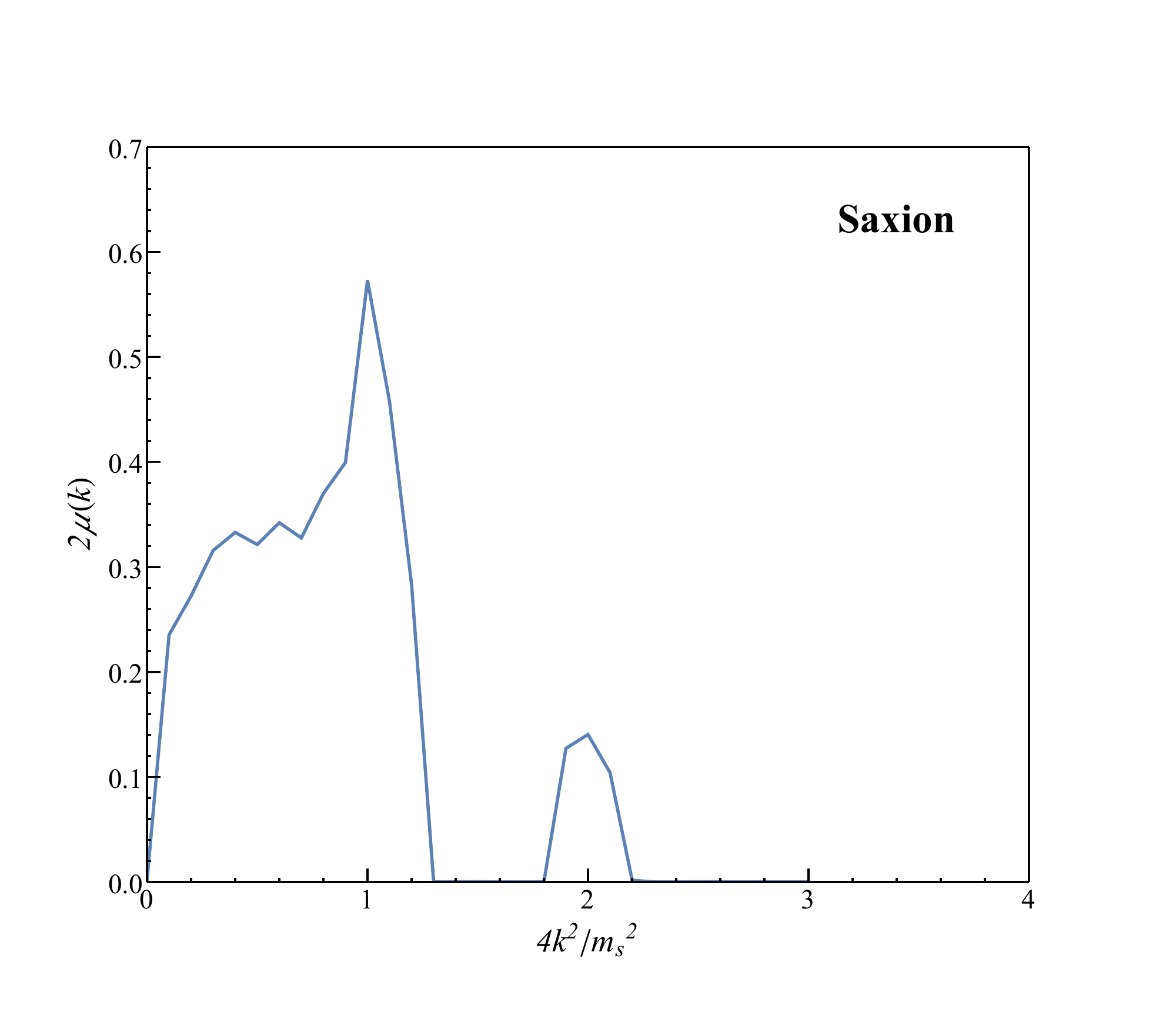}
\caption{Floquet indices for axion (top) and saxion (bottom) modes with a background saxion amplitude of 0.8$f_a$.}
\label{fig:mu}
\end{figure}

In this appendix, we numerically study the process of parametric resonance for our model. To be concrete, let us consider the potential in Eq.~(\ref{eq:V1}) with $n=3$. We decompose the PQ symmetry breaking field as
\begin{align}
P= \frac{1}{\sqrt{2}}\left(f_a+s\left(t\right) + \sigma\left(t,x\right) +i a\left(t,x\right)\right),
\end{align}
where $s$ is a zero mode and $\sigma$ and $a$ are saxion and axion fluctuations, respectively.
The linearized equation of motion of the axion in a mode $k$ is 
\begin{align}
\label{eq:EoM}
\ddot{a}_k + \bigg( k^2 + \frac{m_s^2}{f_a}s + \frac{3m_s^2}{2f_a^2}s^2 +\frac{m_s^2}{f_a^3} s^3 + \frac{m_s^2}{4f_a^4}s^4\bigg)a_k = 0.
\end{align}
For a small saxion oscillation amplitude, a quadratic approximation can be used to give the saxion profile, $s = S_0\sin(m_s t)$, and Eq.~(\ref{eq:EoM}) can be transformed into the Mathieu equation for the axion with a wave number $k$,
\begin{align}
\label{eq:mathieu}
a_k^{\prime\prime}(z)+ \left( A_k-2q\cos\left(2z\right)\right)a_k(z) = 0
\end{align}
where $m_st = 2z-{\pi}/{2}$, $A_k = {4k^2}/{m_s^2}$, and $q= {2S_0}/{f_a}$. For certain values of $(A_k,q)$, the mode solutions exhibit exponential growth of the form $a_k\sim \exp\left(\mu\left(k\right) m_s t\right)$~\cite{Kofman:1997yn}, where $\mu(k)$ is the so-called Floquet index.

Since the amplitude of the saxion oscillations is as large as $f_a$, the quadratic approximation may not be valid and so we numerically solve the saxion zero mode's nonlinear equation of motion, which is given in this case by
\begin{align}
\ddot{s} +\bigg(\frac{m_s^2 }{4 f_a^4}s^4+\frac{5 m_s^2 }{4 f_a^3}s^3+\frac{5 m_s^2 }{2 f_a^2}s^2+\frac{5 m^2_s}{2 f_a}s+m^2_s\bigg) s=0.
\end{align}
The profile determined from this equation of motion is then used to examine the growth of axion modes. This process can also be done for saxion perturbation modes $\sigma_k$, which have the equation of motion
\begin{align}
	\ddot{\sigma}_k + \bigg(k^2 + m_s^2+\frac{5m_s^2}{f_a}s+\frac{15m_s^2}{2f^2_a}s^2+\frac{5m_s^2}{f^3_a}s^3+\frac{5m_s^2}{4f^4_a}s^4\bigg)\sigma_k =0.
\end{align}
The numerical results for the indices $\mu(k)$ for both the axion and saxion are displayed in Fig.~\ref{fig:mu}. The axion and the saxion have similar index profiles and both plots feature a sharp peak at the mode $k \simeq m_s/2$.

\bibliography{LatePQ1}

  \end{document}